\documentclass[final,5p,times,twocolumn]{elsarticle}

\usepackage{bm}
\usepackage{amsmath}
\usepackage{amssymb}
\usepackage{pstricks}
\usepackage{pst-eps}%
\usepackage{pst-plot}%
\usepackage{pst-node}%
\usepackage{pstricks-add}
\usepackage{pst-xkey}
\usepackage{pst-math}
\usepackage{pst-coil}
\usepackage{amssymb}

\renewcommand{\i}{{\rm i}}
\newcommand{\e}{{\rm e}}

\newcommand\ket [1] {|#1 \rangle }

\journal{Physica E}

\begin{document}

\begin{frontmatter}



\title{Effective spin chains for fractional quantum Hall states}


\author[label1]{Emil J. Bergholtz}
\author[label2]{Masaaki Nakamura}
\author[label3]{Juha Suorsa}

\address[label1]{
Max-Planck-Institut f\"{u}r Physik komplexer Systeme,
N\"{o}thnitzer Stra{\ss}e 38, D-01187 Dresden, Germany}
\address[label2]{
Department of Physics, Tokyo Institute of Technology,
Tokyo 152-8551, Japan}
\address[label3]{
Department of Physics, University of Oslo, P.O. Box 1048 Blindern, 0316
Oslo, Norway}

\begin{abstract}

Fractional quantum Hall (FQH) states are topologically ordered which indicates that their essential properties are insensitive to smooth deformations of the manifold on which they are studied. Their microscopic Hamiltonian description, however, strongly depends on geometrical details. Recent work has shown how this dependence can be exploited to generate effective models that are both interesting in their own right and also provide further insight into the quantum Hall system.  We review and expand on recent efforts to understand the FQH system close to the solvable thin-torus limit in terms of effective spin chains. In particular, we clarify how the difference between the bosonic and fermionic FQH states, which is not apparent in the thin-torus limit, can be seen at this level. Additionally, we discuss the relation of the Haldane-Shastry chain to the so-called QH circle limit and comment on its significance to recent entanglement studies.

\end{abstract}


\end{frontmatter}

\section{Introduction}

The quantum Hall (QH) system---cold electrons in two dimensions in a perpendicular magnetic field \cite{qh}---is a striking example of a system where unexpected phenomena emerge at low energies \cite{Laughlin,haldane83,halperin84,jain89,mr}. Ever since its discovery three decades ago, the QH system has inspired a huge amount of experimental and theoretical effort, not least due to its richness in phenomenology and mathematical structure. More recently, it was realized that the theoretical description of a system of rapidly rotating bosons is formally very similar to that of an electron gas in a magnetic field \cite{boseQH}. While in that system the quantum Hall regime (very rapid rotation) is not yet reached experimentally, it provides a further incentive for theorists to consider the QH problem also for bosons. 

Many properties of FQH states are topological in a sense that they are invariant under smooth deformations of the manifold on which we choose to study them. Recent efforts have harnessed this feature and demonstrated that useful insights into the FQH problem can be obtained by studying the interacting many-body problem on the limit geometry of a thin torus, referred to as the the Tao-Thouless (TT) limit \cite{Bergholtz-K2005-8,Seidel-F-L-L-M,ttnonab,Bergholtz-H-H-K} (see also Refs. 
\cite{Tao-T,Anderson,Rezayi-H} for precursory studies). In this proceeding we study the FQH system beyond the TT limit and argue that important characteristics of the states are manifest in the leading quantum fluctuations away from the TT limit. While much of the material presented here has been published earlier \cite{Bergholtz-K2005-8, Wikberg-B-K,Nakamura-B-S,Bergholtz-K2009}, we do present a number of new results. We extend the results of Ref. \cite{Wikberg-B-K} by further establishing that, although the TT limit is largely oblivious to particle statistics, qualitative differences appear in the leading fluctuations. This is consistent with the fact that bosonic and fermionic FQH states are expected at different filling fractions. We also simplify the analysis of an effective spin-$1$ model derived in Ref. \cite{Nakamura-B-S} describing fermions at $\nu=1/3$ by considering an alternative and simpler parameterization thereof. Finally, building on Ref. \cite{Bergholtz-K2009}, we make explicit the connection between the Haldane-Shastry model \cite{hs} and the quantum Hall system in the so-called QH circle limit.
%
\begin{figure}[h]
 \begin{center}
 \includegraphics[width=8.5cm]{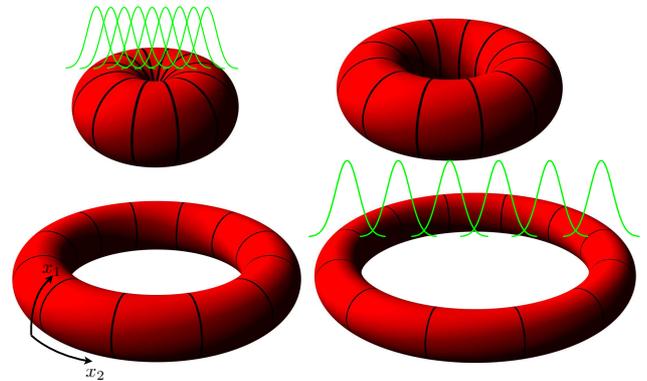}
 \end{center}
 \caption{Relation between the circumference of torus $L_1$ and the distance of
 guiding center coordinates, $2\pi/L_1$. As $L_1$ is decreased, the mutual support of
 neighboring Gaussian-weighted wave functions becomes small, one-dimensional solvable models can be constructed. Also the limit of thick and short cylinder (not shown),
 where the spatial positions of all single particle states coincide, is amenable to
 a detailed analysis.}\label{fig:torus}
\end{figure}

This paper is organized as follows. In Section \ref{map} we describe a mapping of a Landau level problem onto a one-dimensional lattice model and in Section \ref{tt} discuss the TT limit thereof. Section \ref{bulk} gives a number of explicit mappings of the low energy physics beyond the TT limit onto spin chains and offers ideas on how to generalize these special cases. In Section \ref{circle} we turn our attention to the QH circle limit formulated on a short and thick cylinder  \cite{Bergholtz-K2009,Rezayi-H}. Conclusions are given in Section \ref{conclusion}.

\section{Mapping to a one-dimensional model}\label{map}
We consider a model of $N$ interacting particles (fermions or bosons) in the lowest Landau
level on the torus penetrated by $N_\phi$ quanta of magnetic flux. 
In the Landau gauge, a complete basis of $N_\phi$
degenerate single-particle states, labeled by $k=0,\ldots, N_\phi-1$,
can be chosen as
\begin{equation}
\psi_k(x)=
\frac{1}{\sqrt{\pi^{1/2} L_1}}
\sum_{n=-\infty}^{\infty}
\e^{\i (k_1+n L_2) x_1}
\e^{-\tfrac{1}{2}(x_2+k_1+n L_2)^2},
\end{equation}
where $L_i$ are the circumferences of the torus, $x_i$ the corresponding
coordinates, and $k_1=2\pi k/L_1$ the momentum along the $L_1$-cycle.
We have set the magnetic length $l_{\rm B}\equiv \sqrt{\hbar/eB}$ equal
to unity.  In this basis, any translational-invariant 2D two-body
interaction Hamiltonian assumes the form
\begin{equation}
 H=\sum_{k\geq |m|}\hat V_{km},\qquad
  \hat V_{km}\equiv V_{km}
  \sum_{i}
  c_{i+m}^{\dag}
  c_{i+k}^{\dag}
  c_{i+m+k}^{\mathstrut}
  c_{i}^{\mathstrut},
  \label{TT_model}
\end{equation} 
where $c_i$ and $c_i^\dagger$ are either bosonic or fermionic.
The matrix-element $V_{km}$ specifies the amplitude for a pair of particles to hop $m$ sites each towards each other from a separation $k+m$ to a separation
$k-m$ (and vice versa). The boson and fermion matrix elements differ in the sign of the exchange term,
and consequently the amplitude of the
$k=|m|$ process vanishes for a pair of fermions while it does not for bosons.
At the filling $\nu=p/q$ the Hamiltonian commutes with the
center-of-mass magnetic translations $T_1$ and $T_2^q$
along the cycles  \cite{haldane85}, which implies, in particular, that the total momentum
$K$ along the $L_1$-cycle is conserved modulo $N_\phi$ in this gauge.


%

The fermionic Laughlin state at $\nu=1/3$ is an exact zero-energy eigenstate of the 
periodized Haldane pseudo-potential $\nabla^2\delta(\bm{r}-\bm{r}')$ \cite{haldane83,Trugman-K}.
The corresponding hopping amplitudes $V_{km}$ are given by
\begin{equation}
 V_{km}=(k^2-m^2)\e^{-2\left(k^2+m^2\right)\pi^2/L_1^2}\label{ps13}.
\end{equation}
(Strictly speaking, this is the result in the limit $L_2\to \infty$.)
For $\nu\geq 1/3$, this interaction is a good approximation to the Coulomb interaction, which is formally more complicated and includes longer-range components \cite{haldane83}. 

For the boson systems we take the potential as $\delta(\bm{r}-\bm{r}')$
assuming that the boson QH states are realized in trapped Alkali
atoms. Then the hopping matrix elements are given by
\begin{equation}
 V_{km}=\e^{-2\left(k^2+m^2\right)\pi^2/L_1^2}.\label{delta}
\end{equation}
At the level of model wave functions the relation between the bosonic and fermionic states is essentially trivial: multiplying a bosonic state at $\nu=p'/q'$ with the wave function of a filled Landau level gives a fermionic state at filling $\nu=p/q=p'/(q'+p')$. Often, but not always, this provides a simple map between the \emph{ground states} of fermions and bosons. When this works can be understood in terms of pseudopotential considerations.

\section{Thin torus limit of the quantum Hall system}\label{tt}

On the thin torus, the mutual support of different single particle
states vanishes (see Fig. \ref{fig:torus}), and it can be shown that 
the interacting many-body problem becomes solvable for generic interactions. In this limit, the Hamiltonian is diagonal in the occupation number basis and finding the ground state reduces to a classical minimization problem. As an example, the
amplitudes $V_{km}$ in (\ref{ps13}) are exponentially damped in $1/L_1^2$. Therefore, at
small $L_1$ the model can be approximated by a few most dominant terms
such as $\hat V_{10}$, $\hat V_{20}$, {\it etc}. In the case of Coulomb interactions, longer-range electrostatic terms $\hat V_{k0}$ are non-negligible {\it a priori} but turn out to be unimportant. 

Independent of the details of the real-space interaction and the particle statistics, the ground states at any filling fraction are regular lattices, whose mininally charged excitations are domain
walls between degenerate ground states. 
An algebraic expression
for the unit cell is obtained by noting that electron $\alpha$ is at
site 
\begin{equation}\label{uc1} I [\alpha/\nu] \ \ \ , 
\end{equation} 
where $I[x]$ denotes the integer closest to $x$. For example, for $\nu=7/18$, (\ref{uc1}) gives
$\{I[\alpha/\nu]\}=\{0,3,5,8,10,12,15\ldots\}$ and hence the unit
cell is $100101001010010100$. In general, this gives periodic TT ground states with a
unit cell of length $q$ that contains $p$ particles at $\nu=p/q$. Hence, the ground state is $q$-fold
degenerate. This is of course the minimal center-of-mass degeneracy of any state at
$\nu=p/q$ as discussed above. For the abelian
quantum Hall states this is also the topological degeneracy of the ground
state and for these states the TT patterns are  adiabatically connected to the corresponding bulk states \cite{Bergholtz-K2005-8,Bergholtz-H-H-K}.

\section{Beyond the thin torus limit}\label{bulk}

Various scenarios are possible as $L_1$ is increased from zero. Numerical results on small systems indicate that for fermions in the lowest Landau level there is always a phase transition at filling fractions with even denominator but never for odd denominators \cite{Bergholtz-K2005-8}. For bosons this is explored to a lesser extent, but it has been established that at the filling $\nu=1$ a phase transition occurs at finite $L_1$ \cite{Wikberg-B-K}. Below we study the problem by expanding away from the TT limit and explore the properties of the generated effective spin models.

\subsection{Effective spin model for $\nu=1/2$}
For the half-filled Landau level $\nu=1/2$, the electron system on a
torus can be described in terms of $S=1/2$ variables. According to
Ref.~\cite{Bergholtz-K2005-8}, the effective spin
Hamiltonian is obtained in the following way. In the TT limit, the
ground state is the charge-ordered state $|\cdots 010101010\cdots\rangle$, since
the matrix elements of the electrostatic term $\hat{V}_{10}$ are
dominant.  Away from the TT limit, the competition between $\hat V_{10}$, $\hat V_{20}$ and
$\hat V_{2,\pm1}$ can be included as an interaction of
the local spin states $\ket{10}\to\ket{\uparrow}$ and $\ket{01}\to\ket{\downarrow}$. The effective spin
Hamiltonian is then given by $H=\sum_{i=1}^N h_{i,i+1}$ with
\begin{equation}
 h_{ij}=\frac{1}{2}(S_i^+S_j^-+S_i^-S_j^+)+\Delta S_i^zS_j^z,
  \label{XXZ_model}
\end{equation}
where $\Delta=(V_{20}+V_{10})/4V_{21}$. It is well known that in such a XXZ chain, a first-order phase transition from a ferromagnetic (CDW) phase to a gapless phase takes place at $\Delta=-1$.
In the LLL problem, this corresponds to a finite circumference $L_1\simeq
5.3~$ of the torus. This is consistent with the fact that
the $\nu=1/2$ bulk state is gapless \cite{hlr} and there is strong numerical (and circumstantial) evidence that the spin chain solution is smoothly connected to the bulk \cite{Bergholtz-K2005-8}. The result holds for both the local pseudopotential interaction and the Coulomb interaction in the lowest Landau level. However, for the Coulomb interaction in the second Landau level the XXZ spin chain ceases to be a good description. Rather, longer-range electrostatic terms, which appear up as longer-range Ising terms in the spin model, alter the ground state and the low-energy excitations before the hopping terms become non-negligible \cite{Wikberg-B-K}. It can be shown that the new emergent low-energy structure reproduces the level counting and fusion rules of the non-abelian Moore-Read state \cite{mr}, which can alternatively be obtained directly as the TT limit of a system with three-body pseudopotential interactions \cite{ttnonab}.


\subsection{Effective spin-$1/2$ model for $\nu=1$ bosons}
In Ref.~\cite{Wikberg-B-K}, it was shown that a system of $\nu=1$ bosons on a
thin torus can also be described by an effective $S=1/2$ spin chain. As the details of the microscopic mapping are rather involved we only outline the logic here and refer to Ref.~\cite{Wikberg-B-K} for a detailed account. Away from the TT  limit, the total Hilbert space is truncated to a subspace including only the states where any sequence of $n$ consecutive sites contain either $n-1$, $n$ or $n-1$ bosons. This implies approximate charge homogeneity on all length scales and its validity is corroborated by small scale numerical simulations.   
Within this subspace it turns out that the density gradient (with the definition that no change in density give the same spin as the proceeding one) can be used to identify the effective $S=1/2$ degrees of freedom. 
For instance the state $|01112110112\rangle$ map onto the spin state $|\uparrow\uparrow\uparrow\uparrow\downarrow\downarrow\downarrow\uparrow\uparrow\uparrow\downarrow\rangle$ as can be seen by successfully identifying the corresponding density variations in terms of the kinks in the spin state:  $|0^{\uparrow}1^{\uparrow}1^{\uparrow}1^{\uparrow}2^{\downarrow}1^{\downarrow}1^{\downarrow}0^{\uparrow}1^{\uparrow}1^{\uparrow}2^{\downarrow}\rangle$. 
Within this mapping we arrive at a similar effective spin-$1/2$ model as the one obtained for fermions at $\nu=1/2$. However, the effective spin interaction turn out to be slightly different: In contrast to the fermion case, longer-range Ising interactions arise even for the contact interaction (\ref{delta}) in the lowest Landau level, and for rather generic interactions a phase with quasi degeneracies consistent with the non-abelian Moore-Read state is favored \cite{Wikberg-B-K}. This is also consistent with the observation that the Moore-Read state is more stable in the boson system \cite{boseQH}. 

%


\subsection{Effective spin-$1$ model for $\nu=1/3$}

Inspired by the results above, and by the early observation \cite{Girvin-A} of the similarities between the non-local order in FQH states \cite{Girvin-M} and in certain integer $S$ spin chains \cite{Haldane1983,AKLT,Nijs-R}, 
 we recently studied a mapping of the $\nu=1/3$ FQH system onto a spin-$1$ chain. Here we briefly recapitulate this work and provide and alternative and simpler parameterization of the Hamiltonian interpolating between the effective FQH description and conventional Hamiltonians possessing the well studied Haldane \cite{Haldane1983,AKLT} and Large-$D$ phases.
At the filling $\nu=1/3$, the ground state of the truncated Hamiltonian
$\hat V_{10}+\hat V_{20}$ is three-fold degenerate, with the multiplet composed of 
charge-ordered states with one electron in each three-site unit cell: $|\cdots\ 010\
010\ 010\ \cdots\rangle$.  The leading hopping process $\hat V_{21}$ induces fluctuations upon
these ground states through the process
\begin{equation}
 |010\ 010\rangle\leftrightarrow|001\ 100\rangle.
\end{equation}
Again, by identifying the states of the unit cell as $|010\rangle\to|0\rangle$,
$|001\rangle\to|+\rangle$, and $|100\rangle\to|-\rangle$, the truncated
model can be mapped to a $S=1$ quantum spin chain. Validity of this
approximation is confirmed by the high overlap between the
ground state of the full Hamiltonian (\ref{TT_model}), which is
equivalent to the Laughlin state, and that of the truncated Hamiltonian
projected onto the above $S=1$ states par three sites. 

In this mapping, the ground state degeneracy does not have any effect, since the three
degenerate states have different center-of-mass quantum numbers so that there are no
nonvanishing matrix elements between two of these states. Then, in terms of $S=1$
variables, the effective Hamiltonian ${\cal H}=\sum_{i=1}^N h_{i,i+1}$
is given by
\begin{align}
 h_{ij}=&\frac{1}{2}S_i^+S_j^-[1-(S_i^z)^2][1-(S_j^z)^2]+\mbox{H.c.}
 \label{model.0}\\
 =&-\frac{1}{2}S_i^zS_i^+S_j^zS_j^-+\mbox{H.c.}
\end{align}
Note that this Hamiltonian does not have the space inversion
and spin reversal symmetries: the process
$|00\rangle\leftrightarrow|+-\rangle$ exists but
$|00\rangle\leftrightarrow|-+\rangle$ does not.  This parity
breaking is a consequence of the matrix elements $V_{km}$ having different amplitudes
 for inward and outward hopping.
\begin{table}[t]
 \begin{center}
  \begin{tabular}{c|c|rrrrr}
   & & ${\cal P}$ & ${\cal T}$ & $k$& BC & $M$\\ \hline
   $E_0$ & G.S. & $+1$ &$+1$ & $0$ & $+1$ & $0$\\
   $E_1$ & Haldane & $-1$& $-1$& $0$ & $-1$ & $0$\\
   $E_2$ & Large-$D$ & $+1$ & $+1$ & $0$ & $-1$ & $0$\\
   $E_3$ & Dimer & $+1$ &$+1$ & $\pi$ & $-1$ & $0$\\
   $E_4$ & XY & $+1$ & $*$ & $0$ & $+1$ & $2$
  \end{tabular}
 \end{center}
 \caption{Discrete symmetries of the wave functions of the excitation
 spectra (${\cal P}$: space inversion, ${\cal T}$: spin reversal, $k$:
 wave number, and $M$ the total $S^z$). BC$=1$ (BC$=-1$) stands for
 (anti)periodic boundary conditions. G.S. means the ground state. Dimer
 state has wave number $\pi$.}
 \label{tbl:symmetries}
\end{table}
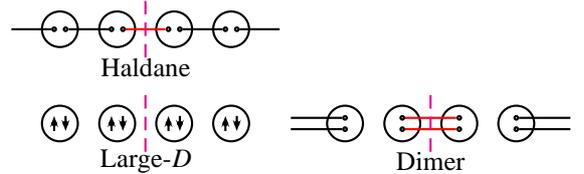
\begin{figure}
\begin{center}
 \psset{unit=2.5mm}
  \begin{pspicture}(0,2)(26,8)
   \rput(0,5){
   \multirput(0,0)(3,0){4}{
   \pscircle(1,2){1}
   \pscircle(0.7,2){0.15}
   \pscircle(1.3,2){0.15}
   }
   \multirput(0,0)(9,0){2}{
   \multirput(0,0)(3,0){2}{
   \psline(-1.55,2)(0.55,2)
   }}
   \psline[linecolor=red,linestyle=solid]{-}(4.45,2)(6.55,2)
   \psline[linecolor=magenta,linestyle=dashed](5.5,0.5)(5.5,3.5)
   \rput[c](5.5,0.0){Haldane}
   }
   \rput(0,0){
   \multirput(0,0)(3,0){4}{
   \pscircle(1,2){1}
   \pscircle(0.7,2){0.15}
   \pscircle(1.3,2){0.15}
   \psline{->}(0.7,1.6)(0.7,2.4)
   \psline{<-}(1.3,1.6)(1.3,2.4)
   }
   \psline[linecolor=magenta,linestyle=dashed](5.5,0.5)(5.5,3.5)
   \rput[c](5.5,0.0){Large-$D$}
   }
   \rput(15,0){
   \multirput(0,0)(3,0){4}{
   \pscircle(1,2){1}
   \pscircle(1,1.7){0.15}
   \pscircle(1,2.3){0.15}
   }
   \multirput(0,0)(12,0){2}{
   \psline(-1.85,2.3)(0.85,2.3)
   \psline(-1.85,1.7)(0.85,1.7)
   }
   \rput(6,0){
   \psline[linecolor=red,linestyle=solid]{-}(-1.85,2.3)(0.85,2.3)
   \psline[linecolor=red,linestyle=solid]{-}(-1.85,1.7)(0.85,1.7)
   }
   \psline[linecolor=magenta,linestyle=dashed](5.5,0.5)(5.5,3.5)
   \rput[c](5.5,0.0){Dimer}
   }
  \end{pspicture}
\end{center} \caption{Valence-bond-solid picture of the three gapped
 states in $S=1$ spin chains with antiperiodic boundary conditions. If
 singlet bonds across the boundary exist, they become triplet
 bonds. This enables us to identify these three different states according
 to discrete symmetry.}  \label{fig:VBS}
\end{figure}

In order to identify the universality class of the ground state of the
model (\ref{model.0}),  we extend the Hamiltonian as
\begin{align}
 h_{ij}=&(1-\lambda)
 \left[
 \bm{S}_i\cdot\bm{S}_j+\frac{D}{2}[(S_i^z)^2+(S_j^z)^2]
 \right]\nonumber\\
 &
 -\lambda
 \left[\frac{1}{2}S_i^zS_i^+S_j^zS_j^-+\mbox{H.c.}\right],
 \label{model}
\end{align}
where the first term is the $S=1$ Heisenberg chain with single ion
anisotropy.  We then study the adiabaticity of deformations from
parameter regions where physical properties are already known, such as
the Haldane phase [$(D,\lambda)=(0,0)$] and large-$D$ phase
[$(D,\lambda)=(2,0)$], to the model which is related to the $\nu=1/3$
FQH state $\lambda=1$. This parameterization provides a simpler alternative to the one studied in Ref. \cite{Nakamura-B-S}.  For $\lambda=0$ it has been known that the
system undergoes a phase transition between the Haldane state ($D<D_{\rm
c}$) and the large-$D$ state ($D>D_{\rm c}$). This critical value is
estimated as $D_{\rm c}=0.968\pm 0.001$. \cite{Chen-H-S2008}

\begin{figure}[t]
\begin{center}
\psset{unit=8mm}
\begin{pspicture}(-1,-0.5)(4,4)
		\psline[linewidth=1.2pt]{->}(0,0)(5,0)
		\psline[linewidth=1.2pt]{->}(0,0)(0,4)
		%
		\psline[linewidth=1.2pt,linecolor=red]{-}(0,3)(3.5,0)
		\psline[linewidth=1.2pt,linecolor=red]{-}(3.5,0)(0,0)
		\psline[linewidth=1.2pt,linecolor=red]{-}(0,0)(0,3)
		\psdot[linecolor=red](0,1.4)
		\rput[c]{0}(5.3,0){$\lambda$}
		\rput[c]{0}(0,4.2){$(1-\lambda)D$}
		\rput[l]{0}(-0.3,0.1){$0$}
		\rput[r]{0}(-0.2,3){$2$}
		\rput[l]{0}(3.6,0.3){$1$}
		\rput[l]{0}(3.5,-0.3)%
		{\textcolor{blue}{$\nu=1/3$}}
		\rput[l]{0}(3.5,-0.8){\textcolor{blue}{FQH}}
		\rput[r]{0}(-0.5,0.5){\textcolor{blue}{Haldane}}
		\rput[r]{0}(-0.5,2.5){\textcolor{blue}{Large $D$}}
		\rput[c]{0}(1.5,-0.4){(a)}
		\rput[c]{0}(1.5,2.1){(b)}
		\rput[r]{0}(-0.2,1.4){$D_{\rm c}\! =\! 0.968$}
\end{pspicture}
\end{center}
\caption{Parameter space of the model (\ref{model}) connecting two
phases (Haldane, large-$D$ phases) in the $S=1$ quantum spin chain
and the $\nu=1/3$ fractional quantum Hall state.}  \label{fig:twodim}
\begin{center}
\includegraphics[height=3.3cm]{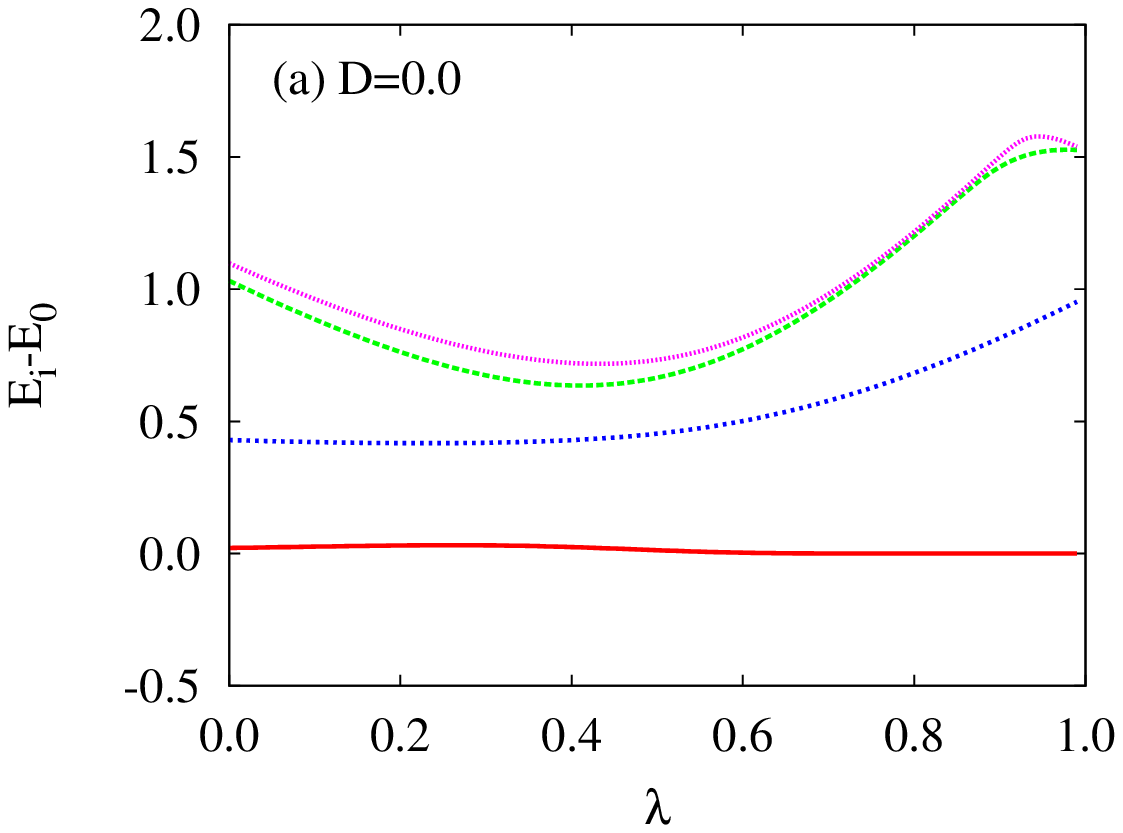}
\includegraphics[height=3.3cm]{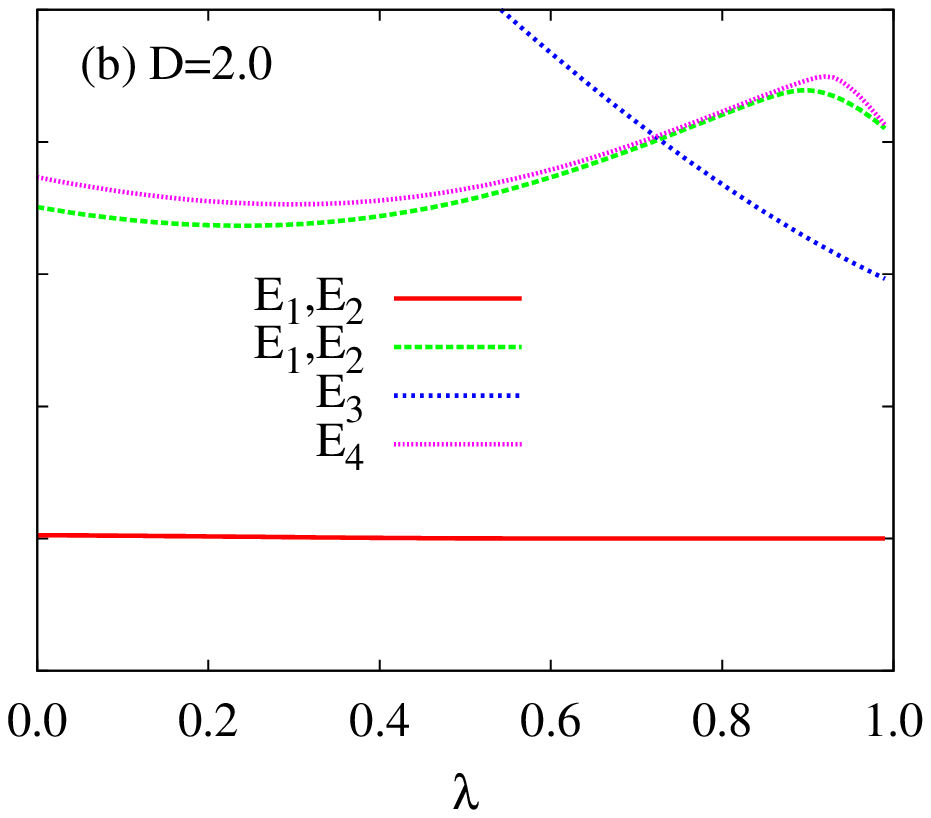}
\end{center}
\caption{Excitation spectra of the $N=18$ system under antiperiodic
boundary conditions for (a) $D=0$ and (b) $D=2$ as functions of
$\lambda$. Although $\lambda=1$ corresponds to the FQH state, and
$(D,\lambda)=(0,0)$ and $(D,\lambda)=(2,0)$ are the Haldane state and
the large-$D$ state, respectively, there are no gap closing points in, since there are no level-crossing points between the lowest
two excitations.}\label{fig:dE_ABC}
\end{figure}

There exists a useful method for analyzing the $S=1$ chain: according to
Refs.~\cite{Nomura-K,Kitazawa}, different phases in $S=1$ chains are
characterized by their spectra of low-lying excitations subject to
twisted boundary conditions, and phase transition points are given by
level crossings of the two lowest energy levels.  These excited states
can be identified according to the discrete symmetry of the wave
functions (summarized in Table.\ref{tbl:symmetries}), as explained by
the valence-bond-solid picture\cite{AKLT} (see Fig.~\ref{fig:VBS}).  In
this analysis, there are four essential excitations that make possible
the identification of the four possible phases. We denote the relevant
energy levels by $E_i$ with $i=0,1,2,3,4$, where $E_0$ is the ground
state energy with periodic boundary conditions.

According to the conventional classification, the gapped state at the
FQH point ($\lambda=1$) would be expected to belong either to the
Haldane or large-$D$ phases. Therefore, we consider the behavior of the
excitation spectra by varying the value of $\lambda$ from $0$ to $1$
with (a) $D=0$ (Haldane to FQH state) and (b) $D=2$ (large-$D$ to FQH
state) as shown in Fig.~\ref{fig:twodim}.  According to the numerical
data of the excitation energies obtained by the exact diagonalization of
$N=18$ clusters, there is no level-crossing point between the lowest two
spectra in these two cases.  This implies that the FQH point is
adiabatically connected to both the Haldane and the large-$D$ phases
(see Fig.~\ref{fig:dE_ABC}).

\begin{figure}[t]
\begin{center}
\psset{unit=5mm}
\begin{pspicture}(0,1)(14,1.5)
\psline[linecolor=red]{->}(0,1)(14,1)
\psdot[linecolor=red](0,1)
\rput(0,1.6){large-$D$}
\rput(7,1.6){large-$D$+Haldane}
\rput(11.5,1.6){?}
\rput(0,0.4){TT state}
\rput(5,0.4){no gap closing}
\rput(11,0.4){Bulk $\nu=1/3$ FQH}
\rput(14.7,1){$L_1$}
\end{pspicture}
\end{center}
 \caption{Phase diagram of the $\nu=1/3$ FQH system as a function of the
 circumference of the torus, $L_1$. The TT limit of the FQH state
 corresponds to the large-$D$ phase of the $S=1$ chain. However, at
 finite $L_1$, large-$D$ and Haldane phases co-exist. For very
 large $L_1$ the FQH/spin-chain correspondence may still be there but this issue is beyond the scope of the present analysis.}\label{fig:pdiagram}
\end{figure}
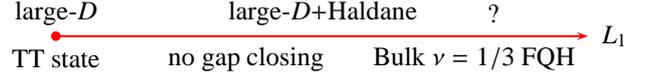

The absence of phase transitions can be understood in terms of the
discrete symmetry of the system. If the system has parity symmetry, the
transition between Haldane and large-$D$ phases indicated by level
crossing of the low-lying excitations \cite{Chen-H-S2008}. However, for
finite $\lambda$, the model lacks parity symmetry, so that this level
crossing is absent.  This is so because there are finite matrix elements
between the two parity sectors which are independent in the
parity-invariant case, and these two energy levels hybridize. Therefore
the absence of the level-crossing is due to the breaking of parity
symmetry.  Arguments for the stability of the Haldane state in terms of
symmetry are also discussed in Refs.~\cite{Berg-T-G-A,Gu-W,Pollmann}.
Thus the FQH state ($\lambda=1$) can be argued to belong to both Haldane
and large-$D$ phases, and we obtain the phase diagram shown in
Fig.~\ref{fig:pdiagram}. We note however, that the FQH state is
described by a dual string order parameter rather than the conventional
one \cite{Nakamura-B-S}. Moreover, the characteristic degeneracies
\cite{Pollmann} of the entanglement spectrum in the symmetry-protected
Haldane phase are absent for the FQH state \cite{Li-H,Lauchli}.

\subsection{Effective spin-$1$ model for $\nu=1/2$ bosons}
Now we show that the $\nu=1/2$ boson system is described by a $S=1$ spin
chain. The leading processes are $\hat{V}_{00}$, $\hat{V}_{10}$ and $\hat{V}_{11}$.
In the TT limit, due to the two electrostatic terms $\hat{V}_{00}$,
$\hat{V}_{10}$, the ground state is the charge-ordered state $|\cdots
010101010\cdots\rangle$.  The hopping term $\hat{V}_{11}$ then
induces the following type of fluctuations:
\begin{equation}
 |\cdots 10101010\cdots\rangle\leftrightarrow
 |\cdots 10020010\cdots\rangle.
\end{equation}
In terms of local $S=1$ spin states $\ket{10}\to \ket{0}$, $\ket{02}\to \ket{+}$ 
and $\ket{00}\to \ket{-}$, this fluctuation process reads
\begin{equation}
 |\cdots 0000\cdots\rangle\leftrightarrow
 |\cdots 0+-0\cdots\rangle.
\end{equation}
Hence, we readily obtain the same effective $S=1$ chain as in the $\nu=1/3$
fermion system given in eq.~(\ref{model.0}). We also note that the obtained $S=1$ description is very robust as the electrostatic and hopping processes collaborate rather than compete. 

\subsection{Effective spin models for generic filling fractions}

Building on the above we can make some comments on the general $\nu=1/q$ case. For fermions the leading hopping terms favor states where two neighboring fermions can be squeezed to be adjacent. In particular this implies that the leading hopping term $V_{21}$ can be activated. While the TT states have this property for odd $q$ this is not the case for even $q$, which leads to a competition between hopping and electrostatics in the latter case. Indeed, when there is competition it seems that the hopping will always wins at some finite $L_1$ leading to a phase transition. In particular, for even $q$, the new ground state will be connected to a unit cell of length $2q$, which has the property that it is hoppable to a state on which the $V_{21}$ terms can act. This scenario is confirmed numerically for small system sizes and $q\leq 11$.

For bosons, the analogous rule is that hopping favors states connected to states on which the $V_{11}$ can act. This leads to the prediction of a phase transition if and only if $q$ is odd. This is indeed the case for $q=1,2$ as discussed above and is likely to generalize to higher $q$.

We also note the adiabatic continuity can in fact be proven in the case of local interactions for $\nu=1/q$ with $q$ odd for fermions and $q$ even for bosons. 

It is tempting to apply an immediate generalization of the above results to general filling fractions $\nu=p/q$ for both fermions and bosons as we will outline below. However, we want to stress that in such a program care is needed and that we have so far not explored to what extent such a generalization is feasible. 
At
this filling, the TT ground states have unit cells of
length $q$ containing $p$ charges as far separated as possible
\cite{Bergholtz-K2005-8}. The $q$ degenerate translations of the unit
cell can be thought of as the $2S+1$ states of a spin $S=(q-1)/2$, which
suggests a mapping of the FQH system at the filling $\nu=p/q$ onto an
effective $S=(q-1)/2$ spin chain. This makes a general connection
between odd (even) denominator fermion FQH fractions and the integer
(half-integer) spin-chains explicit. For bosons this connection is less straightforward,
since the particles need less space just as in the $\nu=1/q$ case.

This is of course very suggestive of the odd denominator rule for abelian
fermion states and the requirement that either the denominator or the
numerator of the filling fraction is even for the bosonic counterparts. 
This is consistent with the recent consistency conditions derived for TT ground states assuming
that they are adiabatically connected to the dual TT limit
\cite{seidel10}, as well as with earlier observations \cite{Tao-W}.

\section{Quantum Hall circle limit and the Haldane-Shastry spin chain}\label{circle}
Another limiting case of the quantum Hall system is the QH circle limit obtained on a cylinder 
by letting the circumference $L_1$ to infinity while keeping the number of sites constant. In this limit the single particle states can be taken as  $\psi_k=\frac 1 {\sqrt{L_1}}\e^{2\pi i kx/L_1}$ and the particles interact via an effective one-dimensional real-space interaction \cite{Bergholtz-K2009,Rezayi-H}. It can be shown that bulk FQH states are adiabatically connected to this limit without closing the gap. For details we refer to Ref. \cite{Bergholtz-K2009} 

There are striking similarities between the 
the Haldane-Shastry (HS) ground state \cite{hs} and the $\nu=1/2$ bosonic Laughlin state.
The former is given by 
\begin{equation}
\ket{\Psi_{\rm
 HS}}=\sum_{i_1,i_2,\ldots,i_K}\psi_{\rm HS}(z_{i_1},\ldots,z_{i_K})
 S^-_{i_1}\cdots S^-_{i_K}\ket{\uparrow_1\cdots \uparrow_{2K}},
\end{equation} 
where 
\begin{equation}\label{eq:hs}
 \psi_{\rm HS}=\prod_{i<j}^K (z_i-z_j)^2\prod_{i}^K z_i\ ;
  \qquad
  z_j=\e^{i \pi j /K}.
\end{equation} 
In the QH circle limit, the $\nu=1/2$ state has essentially the same form:
\begin{equation}\label{eq:L}
\Psi_{\rm L}=\prod_{i<j}^K\sin\left(\frac{\pi}{L_1}(x_i-x_j) \right)^2 = \prod_{i<j}^K (\beta_i-\beta_j)^2\prod_i^K  \beta_i^{-(K-1)},
\end{equation} 
where $\beta_j=\e^{2\pi i x_j/L_1} = \e^{i\pi \hat x_j /K}$, with $\hat x_j = L_2 x_j/2\pi$.
The extra factor $\prod_i^K \beta^{-K}_i$ can be accommodated in (\ref{eq:hs}) or else cancelled in (\ref{eq:L}) by a simple gauge transformation (equivalent to a rigid translation, $T_2^K$). The expression (\ref{eq:L}) is obtained as the
QH circle limit of the 
conformal image of the $\nu=1/2$ state on the disk geometry upon the map $z\rightarrow w = \frac{L_1}{2\pi i}\ln z$, where $z$ and $w$ denote the disk and cylinder coordinates, respectively.

Despite their apparent similarity, the two systems differ in the interpretation of their wave functions. In the HS state the coefficient of a given configuration, with down spins on sites $i_1,i_2,\ldots, i_K$, is (up to an overall normalization common to all configurations) simply given by the value of
$\psi_{\rm HS}(z_{i_1},\ldots,z_{i_K})$ while the route to the occupation number coefficients is
slightly more complicated in the case of the Laughlin state. Here, the coefficients are found by expanding in
the single-particle states. Expanding $\Psi_L$ in terms of the single particle states on the QH circle, one finds 

\begin{eqnarray}
 &\Psi_{\rm L}&=\sum_{m_1,\ldots, m_K} N_{\{m_1,\ldots,m_K\}}\ 
 \beta_{i_1}^{m_1}\cdots \beta_{i_K}^{m_K} \nonumber\\
 &=&(L_1)^{K/2}\sum_{m_1,\ldots, m_K} 
 N_{\{m_1,\ldots,m_K\}}\psi_{m_1}(z_{i_1})\cdots \psi_{m_K}(z_{i_K}),
\end{eqnarray}
where the symmetric array $N_{\{m_1,\ldots,m_K\}}$ consists of integers giving the
occupation number coefficients. If the same expansion of the (full bulk) Laughlin state is done in Landau gauge on
the cylinder one finds that the coefficients are given by
\begin{equation}
N_{\{m_1,\ldots,m_K\}}\rightarrow\e^{2\pi^2\sum_i^K m_i^2/L_1^2}N_{\{m_1,\ldots,m_K\}}
\end{equation} which smoothly approaches the above expression in the circle limit, $L_1\rightarrow\infty$.

In Ref. \cite{ronny} the ``sites'' in the HS state
are reinterpreted by expanding in a ``momentum'' basis,
$\tilde{\psi}_m=z^m$ \cite{note}. By construction this will give exactly the same expansion coefficients as in the Laughlin state in the circle limit.  Thus, this directly implies that the counting of entanglement levels \cite{Li-H} in the HS and Laughlin states are the same which is indeed concluded on the basis of numerical simulations in Ref. \cite{ronny}. Moreover, the above analysis clearly shows that the entanglement spectra are identical for the Haldane-Shastry ground state and the Laughlin state in the QH circle limit and that one can continuously interpolate this spectra to that of the bulk Laughlin state by varying $L_1$. In this context we note that, it was recently shown that no qualitative changes occur in the entanglement spectra as $L_1$ is varied, even in the opposite limit $L_1\rightarrow 0$ \cite{Lauchli} (the von Neumann entanglement entropy is however dependent on $L_1$ and vanishes as $L_1\rightarrow 0$ \cite{deform}). Moreover, we note that the $L_1\rightarrow\infty$ limit on the cylinder makes the tentative notion of "conformal limit" recently introduced in Ref. \cite{ronny2} explicit.

%



\section{Conclusions}\label{conclusion}

We have studied the quantum Hall system of fermions as well as bosons taking into account the leading corrections (quantum fluctuations) beyond the Tao-Thouless limit. This naturally lead to the study of effective spin models, within which we have addressed the issue of which TT states are robust and which are not. Explicitly, we have seen that the TT patters melt in the case of fermions at $\nu=1/2$ and bosons at $\nu=1$ where the system naturally maps onto spin-1/2 chains. In contrast, the $\nu=1/3$ fermion and $\nu=1/2$ boson system map onto a spin-1 model whose dynamics cooperate with the TT state in a way that the charge order prevails and the gap remains finite. More generally this allows us to speculate on a possible more general connection between the properties of spin $S=(q-1)/2$ spin chains and the quantum Hall system at filling fraction $\nu=p/q$ for fermions and  $\nu=p/(q-p)$ for bosons.

The obtained spin-models are also interesting in their own right, and in some cases possess rather unconventional properties. In particular this is the case for the $S=1$ chains which break parity and have ground states characterized by coexisting features of the large-$D$ and Haldane phases, that would be separated by a phase boundary if the parity symmetry would not be broken. 



We have also made the connection between the Haldane-Shastry model and the quantum Hall effect explicit by considering the QH circle limit, and discussed the significance of this connection to recent entanglement studies.

\section*{Acknowledgments} 
E. J. B.  acknowledges Anders Karlhede and Emma Wikberg for related collaborations and
discussions.  M.~N. thanks Zheng-Yuan Wang for discussions.
M.~N. acknowledges support from Global Center of Excellence Program
``Nanoscience and Quantum Physics'' of the Tokyo Institute of Technology
by MEXT.





\bibliographystyle{elsarticle-num}
\bibliography{<your-bib-database>}

\begin{thebibliography}{00}

\bibitem{qh} K.v. Klitzing, G. Dorda and M. Pepper, Phys. Rev. Lett. 
{\bf 45}, 494 (1980); D.C. Tsui, H.L. Stormer and A.C. Gossard, Phys. Rev. Lett.
{\bf 48}, 1559 (1982).

 \bibitem{Laughlin}
	 R.~B.~Laughlin,
	 Phys. Rev. Lett. {\bf 50}, 1395 (1983).

 \bibitem{haldane83}
	 F. D. M.  Haldane, Phys. Rev. Lett. {\bf 51}, 605 (1983).

 \bibitem{halperin84}
	 B. I.  Halperin,  Phys. Rev. Lett. {\bf 52}, 1583 (1984). 

 \bibitem{jain89}
	J. K. Jain, Phys. Rev. Lett. {\bf 63}, 199 (1989).

 \bibitem{mr}
	 G. Moore and N. Read, 
	 Nucl. Phys. B {\bf 360}, 362 (1991);  N. Read and E.H. Rezayi,
	 Phys. Rev. B {\bf 59}, 8084 (1999).

 \bibitem{boseQH}
	 N.K. Wilkin, J.M.F. Gunn and R.A. Smith, Phys. Rev. Lett. {\bf 80},  2265  (1998); N.R. Cooper and N.K. Wilkin, Phys Rev. B {\bf 60} (1999) 16279(R); N. Regnault and Th. Jolicoeur, Phys. Rev. Lett. {\bf 91},  030402 (2003); For a recent review see S. Viefers J. Phys.: Condens. Matter {\bf 20}, 123202 (2008).


 \bibitem{Bergholtz-K2005-8}
	 E.~J.~Bergholtz and A.~Karlhede, Phys. Rev. Lett. {\bf 94}, 026802 (2005);
	 J. Stat. Mech. (2006) L04001 [arXiv:cond-mat/0509434];
	 Phys. Rev. B {\bf 77}, 155308 (2008).

 \bibitem{Seidel-F-L-L-M}
	 A.~Seidel, H.~Fu, D.~-H.~Lee, J.~M.~Leinaas and J.~Moore,
	 Phys. Rev. Lett. {\bf 95},  266405 (2005).

 \bibitem{ttnonab}
	 E.J. Bergholtz, J. Kailasvuori, E. Wikberg, T.H. Hansson and
	 A. Karlhede, Phys. Rev. B {\bf 74}, 081308(R) (2006); A. Seidel and
	 D.-H. Lee, Phys. Rev. Lett. {\bf 97}, 056804 (2006); N. Read,
	 Phys. Rev. B {\bf 73},  245334 (2006); E. Ardonne, E.J. Bergholtz,
	 J. Kailasvuori and E. Wikberg, J. Stat. Mech. (2008) P04016 [arXiv:0802.0675].

 \bibitem{Bergholtz-H-H-K}
	 E.~J.~Bergholtz, T.~H.~Hansson, M.~Hermanns and A.~Karlhede,
	 Phys. Rev. Lett. {\bf 99}, 256803 (2007).

 \bibitem{Tao-T}
	 R.~Tao and D.~J.~Thouless,
	 Phys. Rev. B {\bf 28},  1142 (1983).

 \bibitem{Anderson}
	 P.~W.~Anderson,
	 Phys. Rev. B {\bf 28},  2264 (1983).

 \bibitem{Rezayi-H}
	 E.~H.~Rezayi and F.~D.~M.~Haldane,
	 Phys. Rev. B {\bf 50}, 17199 (1994).
	 

 \bibitem{Wikberg-B-K}
	 E.~Wikberg, E.~J.~Bergholtz and A.~Karlhede,
	 J. Stat. Mech. (2009) P07038 [arXiv:0903.4093].

 \bibitem{Nakamura-B-S}
	 M. Nakamura, E. J. Bergholtz and J. Suorsa,
	 Phys. Rev. B {\bf 81}, 165102 (2010).
	

 \bibitem{Bergholtz-K2009}
	 E.~J.~Bergholtz and A.~Karlhede,	 
	 J. Stat. Mech. (2009) P04015 [arXiv:0902.0167].
	 

 \bibitem{hs}
	 F.D.M. Haldane,   Phys. Rev. Lett. {\bf 60}, 635 (1988); B.S. Shastry,
	 Phys. Rev. Lett. {\bf 60}, 639 (1988). For a pedagogical account, see
	 also M. Greiter, J. Low Temp. Phys. {\bf 126} 1029 (2002).		 


	
 \bibitem{haldane85}
	 F.~D.~M.~Haldane,
	 Phys. Rev. B. {\bf  31}, 2529 (1985).

 \bibitem{Trugman-K}
	 S.~A.~Trugman and S.~A.~Kivelson,
	 Phys. Rev. B {\bf 31}, 5280 (1985).
	 
	 

	 \bibitem{hlr} B.I. Halperin, P.A. Lee, and N. Read, Phys. Rev. B {\bf 47}, 7312 (1993); E.H.  Rezayi, and N. Read,  Phys. Rev. Lett. {\bf 72}, 900 (1994).
	 	 


 \bibitem{Girvin-A}
	 S.~M.~Girvin and D.~P.~Arovas,
	 Physica Scripta. T 27 (1989) 156-159.

 \bibitem{Girvin-M}
	 S.~M.~Girvin and A.~H.~MacDonald,
	 Phys. Rev. Lett. {\bf  58}, 1252 (1987).


 \bibitem{Haldane1983}
	 F.~D.~M.~Haldane,
	 Phys. Lett. 93A, 464 (1983);
	 Phys. Rev. Lett. {\bf  50}, 1153 (1983). 


 \bibitem{AKLT}
	 I. Affleck, T. Kennedy, E. Lieb and H. Tasaki,
	 Phys. Rev. Lett. {\bf 59}, 799 (1987);
	 Commun. Math. Phys. {\bf 115}, 477 (1988).

 \bibitem{Nijs-R}
	 M.~den Nijs and K.~Rommelse, Phys. Rev. B {\bf  40}, 4709 (1989).

	 

 \bibitem{Chen-H-S2008}
 W.~Chen, K.~Hida and B.~C.~Sanctuary, Phys. Rev. B {\bf 67}, 104401 (2003); J. Phys. Soc. Jpn. {\bf  77},  118001 (2008).

 \bibitem{Nomura-K}
	 K.~Nomura and A.~Kitazawa,
	 J. Phys. A: Math. Gen. {\bf  31}, 7341 (1998).

 \bibitem{Kitazawa}
	 A. Kitazawa, J. Phys. A {\bf  30} (1997) L285.


 \bibitem{Berg-T-G-A}
	 E. Berg, E. G. Dalla Torre, T. Giamarchi and E. Altman,
	 Phys. Rev. B {\bf 77},  245119 (2008). 
	 See also E. G. Dalla Torre, E. Berg, E. Altman,
	 Phys. Rev. Lett. {\bf  97}, 260401 (2006).

 \bibitem{Gu-W}
	 Z.-C. Gu and X.-G. Wen, Phys. Rev. B {\bf  80}, 155131 (2009).
	 
 \bibitem{Pollmann} 
	 F. Pollmann, E. Berg, A. Turner, and M. Oshikawa,
	 arXiv:0909.4059 (unpublished); F. Pollmann, A. M. Turner,
	 E. Berg, and M. Oshikawa, Phys. Rev. B {\bf  81},  064439 (2010).

 \bibitem{Li-H}
	 H. Li and F.D.M.~Haldane, Phys. Rev. Lett. {\bf 101}, 010504 (2008).


 \bibitem{Lauchli}
	 A.~M. L\"auchli,  E.~J. Bergholtz, J. Suorsa and M. Haque,
	 Phys. Rev. Lett. {\bf 104}, 156404 (2010).
 

	 
	 \bibitem{seidel10} A. Seidel, Phys. Rev. Lett. {\bf 105},  026802  (2010).
	 
	 \bibitem{Tao-W} R. Tao and Y.-S. Wu, Phys. Rev. B {\bf  31},  6859 (1985).
 
	 
 \bibitem{ronny}
	 R. Thomale, D. P. Arovas, B. A. Bernevig.  arXiv:0912.0028.
	 
\bibitem{deform}
	 A.~M. L\"auchli,  E.~J. Bergholtz, and M. Haque, New J. Phys. {\bf 12},  075004 (2010).
	
	 
 \bibitem{ronny2} R. Thomale, A. Sterdyniak, N. Regnault, B. A. Bernevig, Phys. Rev. Lett. {\bf 104},  180502 (2010).
	 
	 
 \bibitem{note} Note, however, that the real-space hard-core constrain fulfilled by the spin operators become very complicated in momentum space. Thus it is not a priori clear to what extent such a procedure is useful.
	 
\end{thebibliography}

\end{document}